\documentclass[12pt,dvips,a4paper]{article}
\usepackage{graphicx,a4,epsf,here,exscale}
\newcommand{\Figurebb}[9]{
 \begin{figure}[H]\begin{center}
 \leavevmode
 \epsfysize=#7cm
 \epsfbox[#2 #3 #4 #5]{#6}
 \par
 \parbox{#8cm}{
 \caption[figure]{\renewcommand{\baselinestretch}{0.8} \small 
                  \hspace{-0.3truecm}#9}
 \label{#1}
}
\end{center}
\end{figure}}
\def\fig#1{Fig.~\ref{#1}}
\def\bs{\bigskip}
\def\ms{\medskip}
\def\eq#1{(\ref{#1})}

\def\Cosh{\hbox{Cosh}}
\def\be{\begin{equation}}
\def\ee{\end{equation}}
\def\etal{{\it et al.}}
\begin{document}

\baselineskip 16pt

\vglue -1.4truecm

\hfill TPR-00-08

\bs

\centerline{\bf BIFURCATION CASCADES AND SELF-SIMILARITY OF}

\bs

\centerline{\bf PERIODIC ORBITS WITH ANALYTICAL SCALING CONSTANTS}

\bs

\centerline{\bf IN HENON-HEILES TYPE POTENTIALS}

\bs
\bs

\centerline{\bf Matthias Brack}

\ms
\bs

\centerline{\it Institute for Theoretical Physics, University of
Regensburg}

\centerline{\it D-93040 Regensburg, Germany}

\bs
\bs

\centerline{\bf Abstract}

\begin{center}
\begin{minipage}{13.5cm}
\baselineskip 10pt
\small{
We investigate the isochronous bifurcations of the straight-line
librating orbit in the H\'enon-Heiles and related potentials. With
increasing scaled energy $e$, they form a cascade of pitchfork
bifurcations that cumulate at the critical saddle-point energy $e=1$.
The stable and unstable orbits created at these bifurcations appear in
two sequences whose self-similar properties possess an analytical
scaling behavior. Different from the standard Feigenbaum scenario in
area preserving two-dimensional maps, here the scaling constants
$\alpha$ and $\beta$ corresponding to the two spatial directions are
identical and equal to the root of the scaling constant $\delta$ that
describes the geometric progression of bifurcation energies $e_n$ in
the limit $n\rightarrow\infty$. The value of $\delta$ is given
analytically in terms of the potential parameters.
}
\end{minipage}
\end{center}

\bs
\bs

\baselineskip 16pt

{\bf 1. Introduction}

\ms

\noindent
The present study arose in the context of applying Martin Gutzwiller's
semiclassical trace formula \cite{gutz,gubu}, and some of its
extensions, to various model Hamiltonians and interacting fermion
systems in the mean-field approximation, with the aim of describing
prominent quantum shell effects semiclassically in terms of the leading
classical periodic orbits with shortest periods. This approach was
promoted by Strutinsky \etal\ \cite{stru}, who estimated the nuc\-lear
ground-state deformations from the shortest periodic orbits in an
ellipsoidal billiard. Applications which involved the author were beats
in the level density of the H\'enon-Heiles and related potentials
\cite{hh1,hhun}, conductance oscillations in mesoscopic semiconductor
structures \cite{qdot,jo}, and the onset of mass asymmetry in nuclear
fission \cite{fiss}. We refer to the literature just quoted and to a
recent monograph \cite{book} for a detailed discussion of the
extensions of Gutzwiller's theory that are adequate for treating the
degenerate orbit families occurring in systems with continuous
symmetries. Uniform approximations that become necessary in connection
with symmetry breaking and bifurcations will be referred to in the next
section. The accumulated experience from these investigations is an
astonishing performance of the periodic orbit theory in reproducing
quantum-mechanical gross-shell structure, using just a few short 
periodic orbits in the appropriate semiclassical trace formulae.

In the present paper, we will stay on a purely classical level and
investigate the onset of chaos in the H\'enon-Heiles and related
potentials through bifurcation cascades, the accompanying 
self-similarity of periodic orbits, and their analytical scaling 
behavior.

%\bs
\newpage

{\bf 2. The H\'enon-Heiles potential}

\ms

\noindent
We investigate here the role of the straight-line librating orbit A in
the H\'enon-Heiles (HH) Hamiltonian \cite{hh}:
\begin{equation}
H = \frac{1}{2}\,({\dot x}^2+{\dot y}^2) + \frac{1}{2}\,(x^2+y^2)
  + \varepsilon \, (x^2y-\frac13\,y^3) \,.
\label{hhxy}
\end{equation}
Introducing the scaled variables $u=\varepsilon x$ and $v=\varepsilon
y$, the scaled total energy $e$, in units of the saddle-point energy
$E^*=1/6\varepsilon^2$, becomes
\begin{equation}
e = E/E^* = 6\left[\frac12 ({\dot u}^2 + {\dot v}^2) + V(u,v)\right]
  = \,3\,({\dot u}^2+{\dot v}^2)+3\,(u^2+v^2)+6\,v\,u^2-2\,v^3.
\label{scale}
\end{equation}
The Newton equations of motion in $u,v$ are
\begin{eqnarray}
\ddot u & = & - u\,(1 + 2v)\, , \nonumber \\
\ddot v & = & - v + v^2 - u^2 .
\label{hheom}
\end{eqnarray}
These equations, and therefore the classical dynamics of the HH
potential, depend only on the scaled energy $e$ as a single parameter.
For our numerical investigations below, we have solved Eqs.\
(\ref{hheom}) numerically and determined the periodic orbits by a
Newton-Raphson iteration using their stability matrix \cite{kabr}.

In the left part of \fig{hhcont}, we show the equipotential lines in
the $(u,v)$ plane. The lines for $e=1$ intersect at the saddle points
and form an equilateral triangle. Along the three symmetry axes (dashed
lines) the potential is a cubic parabola as shown, e.g., along $u=0$ in
the right-hand part of \fig{hhcont}. The figure also shows the three
shortest periodic orbits.

\Figurebb{hhcont}{-44}{261}{649}{561}{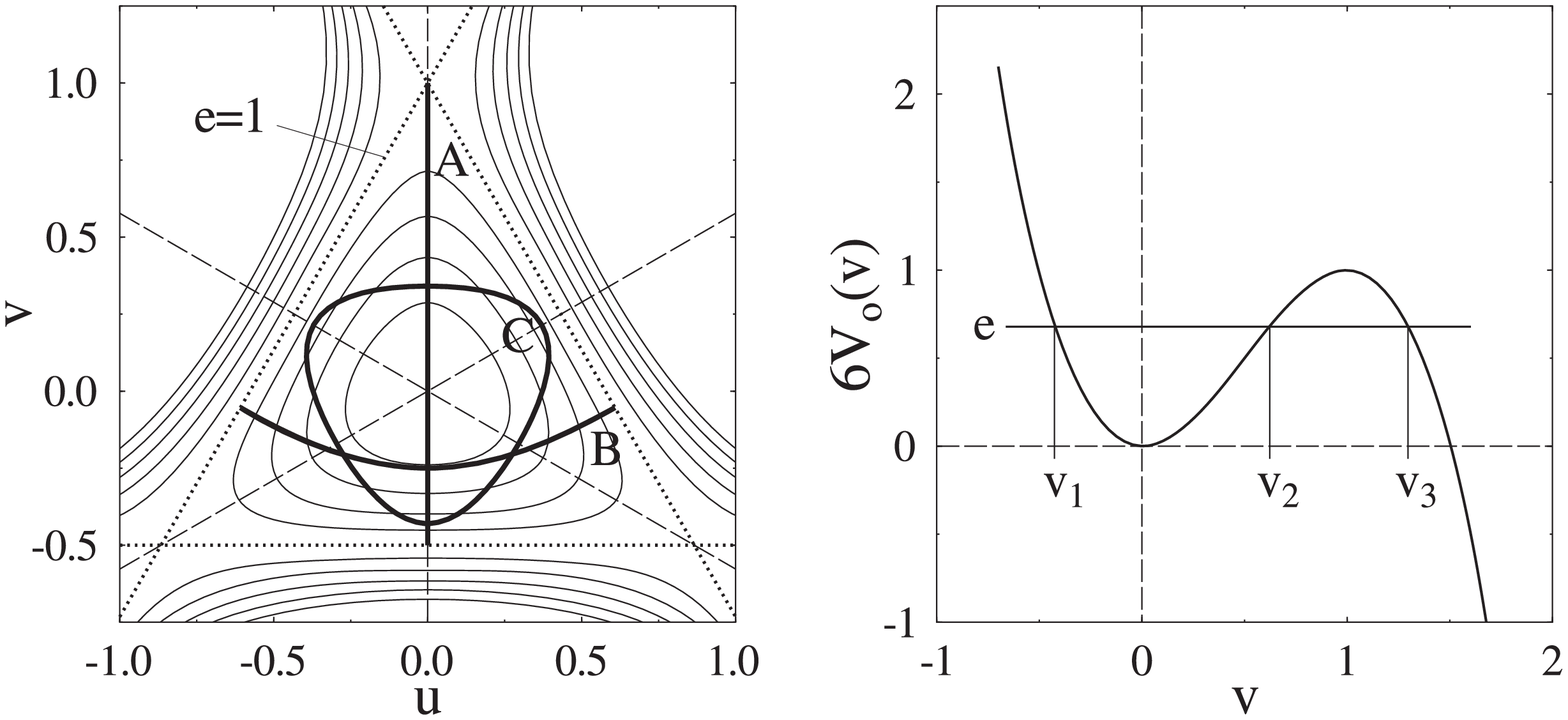}{5}{14.5}{
The H\'enon-Heiles potential. {\it Left side:} equipotential lines
in the $(u,v)$  plane. The dashed lines are the symmetry axes. The
three shortest periodic orbits A, B, and C (evaluated at $e=1$) are
shown by the heavy solid lines. {\it Right side:} scaled potential
along $u=0$. (After \cite{hh1}.)
}

H\'enon and Heiles \cite{hh} have already observed that the classical
motion in this potential is quasi-regular up to energies $e\sim 0.5$
and then becomes increasingly chaotic; when one reaches the saddle
energy ($e=1$), more than $95\%$ of the phase space is covered
ergodically. The periodic orbits in the HH potential have been
classified and investigated in detail by Churchill \etal\ \cite{chur}
and more recently by Davies \etal\ \cite{hhdb}. Up to $e\simeq 0.97$
there exist only three types of periodic orbits with periods of the
order of $T_0=2\pi$ (i.e., the fundamental period of the
harmonic-oscillator potential reached in the limit $e\rightarrow0$):
the librations A and B, and the rotation C. Corresponding to the
symmetry of the HH potential, orbits A and B occur in three
orientations connected by rotations in the $(u,v)$ plane about $2\pi/3$
and $4\pi/3$. Orbit C maps unto itself under these rotations but has
two opposite time orientations, whereas time reversal maps the orbits A
and B onto themselves. The overall discrete degeneracies are thus three
for orbits A and B, and two for orbit C. The orbits A are stable up to
$e\simeq 0.81$ where they become unstable at a period-doubling
bifurcation. At higher energies they oscillate between stability and
instability, undergoing an infinite number of isochronous bifurcations
that cumulate at the saddle-point energy $e=1$ where the period $T_A$
becomes infinity. This bifurcation cascade is the main object of our
present study. For $e>1$ the orbits A do not exist any more, but a new
set of three degenerate unstable librations across the saddle points
come into being \cite{chur,hhdb,vioz}; in accordance with Ref.\
\cite{vioz} we call them $\tau$. The orbits B are unstable at all
energies; the orbits C are stable at low energies and undergo a
period-doubling bifurcation at the energy $e\simeq 0.89$ beyond which
they remain unstable. A, B, and C are the only generic periodic orbits
in the HH system for $e \leq 1$; all other orbits are created through
their bifurcations (and further sequential bifurcations).

\Figurebb{hhdg}{75}{60}{759}{555}{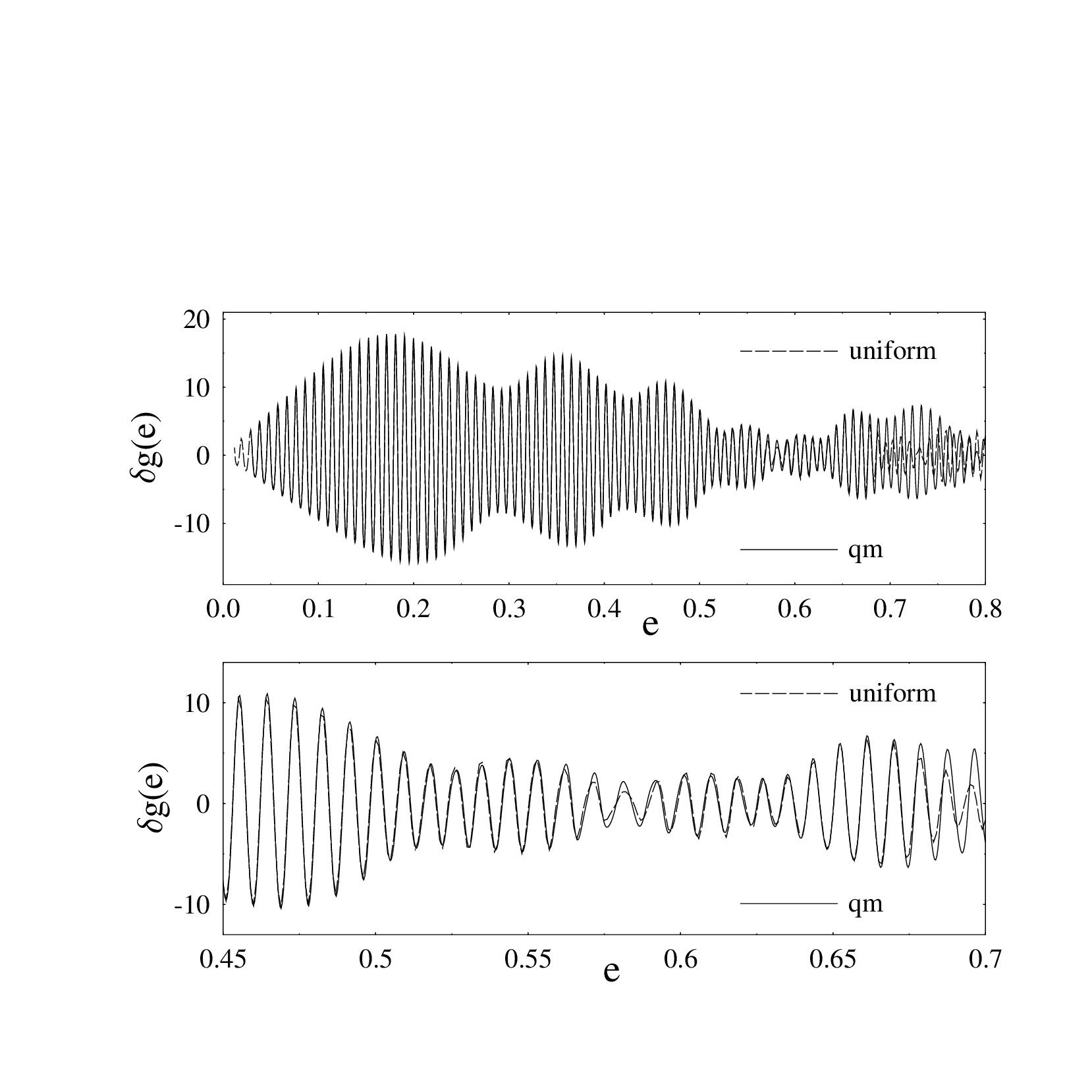}{9}{14.5}{
Oscillating part of level density of the HH potential, Gaussian
averaged over an energy range $\Delta e=0.0024$, versus scaled energy
$e$. {\it Solid line:} quantum-mechanical result (evaluated for
$\varepsilon=0.04$). {\it Dashed line:} semiclassical result, obtained
with the uniform trace formula of Ref.\ \cite{hhun} including first and
second repetitions of the three orbits A, B, and C.
}

In Ref.\ \cite{hh1} the coarse-grained quantum level density of the HH
potential was shown to exhibit a pronounced beating structure which
could be reproduced quantitatively by Gutzwiller's semiclassical trace
formula \cite{gutz} including the orbits A, B, and C. In the limit
$e\rightarrow 0$, the trace formula diverges due to the approaching
harmonic-oscillator limit which is integrable and has SU(2) symmetry
(and a continuous two-fold degeneracy of its periodic orbits). This
divergence can be removed in a uniform approximation that has recently
been derived for some specific cases of SU(2) and SO(3) symmetry
breaking \cite{hhun}. In \fig{hhdg} we show a comparison of the
oscillating part $\delta g(e)$ of the quantum-mechanical level density
$\delta g(e)$ of the HH potential with its semiclassical approximation
obtained from the uniform trace formula of Ref.\ \cite{hhun} including
only the first and second repetitions of the three orbits A, B, and C.
Both results have been coarse-grained by convoluting them with a
Gaussian over an energy range $\Delta e=0.0024$. The agreement is
excellent up to $e \simeq 0.67$. The errors at higher energies are
expected to come mainly from the orbit bifurcations that have not been 
taken into account (and partly from inaccuracies in the 
quantum result \cite{hh1,hhun}).

An attempt to include the bifurcations in the semiclassical approach
has lead to our present study. Uniform approximations for isolated
bifurcations of all generic types have been developed by Sieber and
Schomerus \cite{ssun} and successfully applied to the semiclassical
description of various systems with mixed dynamics (see also Ref.\
\cite{mawu} for an alternative treatment of the three simplest
bifurcation types). Interferences of two close-lying bifurcations
(so-called bifurcations of codimension two) were discussed in Ref.\
\cite{scho}. However, none of these uniform approaches can be used in
the present case of the orbit A, where an infinite number of
bifurcations coalesce at the saddle energy $e=1$. The uniform treatment
of this bifurcation cascade in a semiclassical trace formula is a
challenging task. We should mention that an infinity of orbit
bifurcations in the integrable two-dimensional elliptic billiard has
recently been incorporated successfully into an analytical trace
formula \cite{elli}. The bifurcating orbit in the ellipse -- the
straight-line libration along the shorter diameter -- and the orbit
families created at its bifurcations are, however, of a rather simple
nature, and it is not clear yet if we can apply the technique of Ref.\
\cite{elli} to the present system. While working along this line
\cite{safe}, it seemed worth while to investigate on a purely classical
level the bifurcation sequence of the saddle orbit A in the HH and
similar potentials, which bears a lot of resemblance to the famous
Feigenbaum scenario observed in one-dimensional \cite{feig} and
two-dimensional maps \cite{fei2,fei3}. We shall presently exhibit the
self-similarity amongst the periodic orbits created at the successive
bifurcations and show that it is quantitatively described by the same
scaling constant $\delta$ that accounts for the geometric progression
of the bifurcation energies. Different from the Feigenbaum scenario,
the constant $\delta$ is given analytically here, but it is not
universal in that it depends on the parameters of the potential.

A convenient method to keep track of orbit bifurcations in a 
two-dimensional system is to plot the
trace of their stability matrix, tr\,M. Bifurcations occur whenever
tr\,M $=2$. In Fig.\ \ref{zoom} we show tr\,M for the primitive orbit
A$_n$ (with its Maslov index $n$ increasing by one unit at each
bifurcation) and for the orbits born at its bifurcations. In the lowest
panel, we see the uppermost 3\% of the energy scale available for the
orbit A. The first bi-

\Figurebb{zoom}{-52}{41}{656}{780}{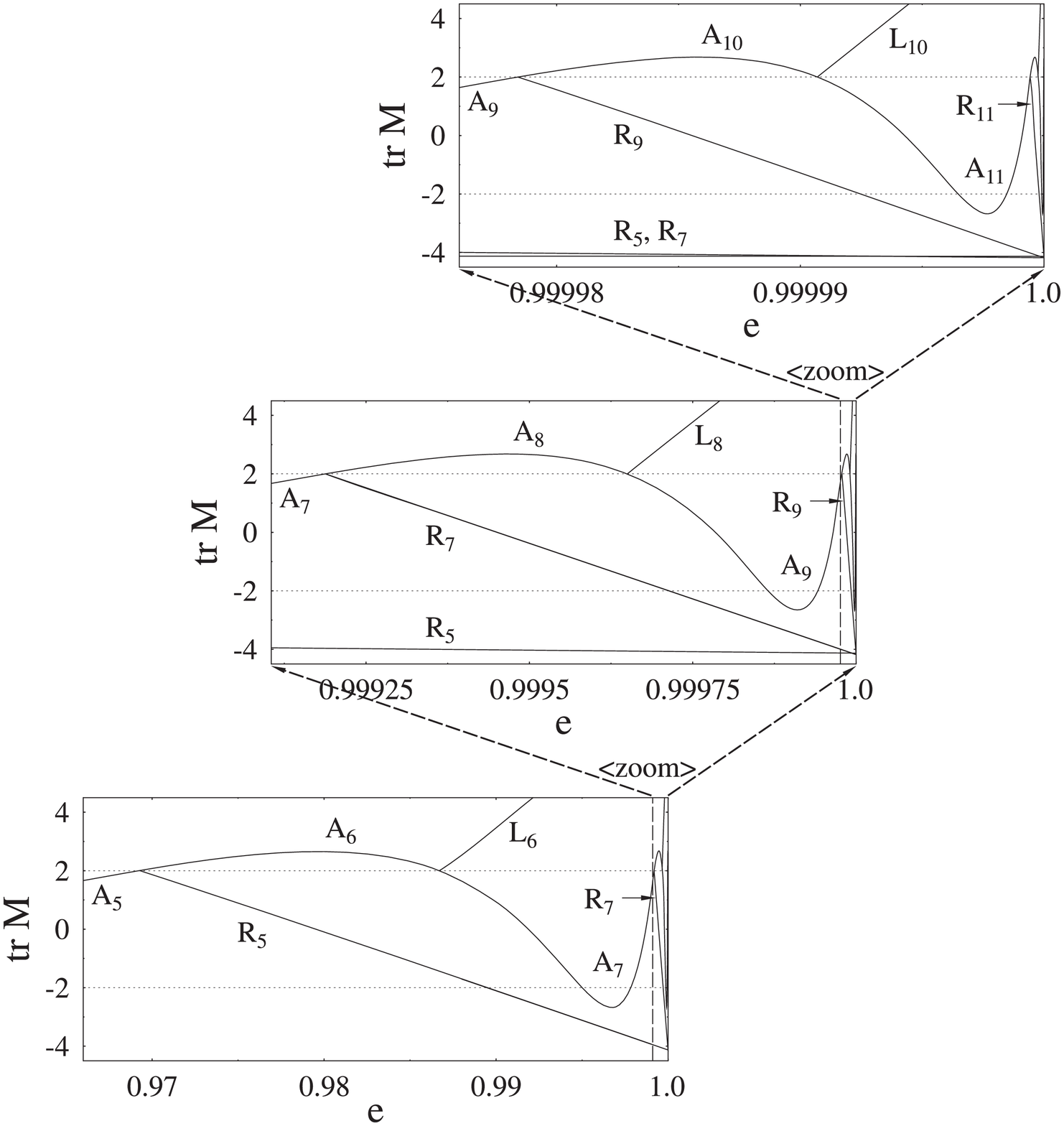}{13.5}{14.5}{
Trace of stability matrix M of orbit A and the orbits born at
successive pitchfork bifurcations in the H\'enon-Heiles potential,
versus scaled energy $e$. Subscripts are Maslov indices. {\it From
bottom to top:} successively zoomed energy scale near $e=1$.
}

\vspace*{-0.25cm}

\noindent
furcation occurs at $e_5 = 0.969309$, where A$_5$
becomes unstable (with tr\,M$_A$ $>2$) and a new stable orbit R$_5$ is
born. At $e_6 = 0.986709$, orbit A$_6$ becomes stable again and a new
unstable orbit L$_6$ is born. In the middle panel, we have zoomed the
uppermost 3\% of the previous energy scale. Here the behavior of A
repeats itself, with the new orbits R$_7$ and L$_8$ born at the next
two bifurcations. Zooming with the same factor to the top panel, we see
the birth of R$_9$ and L$_{10}$. This can be repeated {\it ad
infinitum}: each new figure will be a replica of the previous one, with
all the Maslov indices increased by two units and with tr\,M$_A$
oscillating forever.

Note that we have only shown here the primitives (i.e., the first
repetitions) of each orbit. The higher repetitions of A will also
undergo regular bifurcations and exhibit a corresponding fractal
behavior. (For instance, whenever tr\,M of an orbit becomes equal to
$-2$, its second repetition will have tr\,M $=2$ and bifurcate.) This
infinite proliferation of stable and unstable orbits creates an
increasingly mixed phase space and paves the way to chaos, similarly to
the well-known Feigenbaum scenario.

We should emphasize an important difference, though, to the Feigenbaum
scenario of Refs.\ \cite{feig,fei2,fei3}: the bifurcations investigated
there were all period doublings. Following the new stable orbit born at
each bifurcation to its next period-doubling bifurcation leads to the
famous Feigenbaum tree with its fractal structure. In our present
system, however, successive pitchfork bifurcations occur from one and
the same orbit A. Due to the discrete symmetries of the HH potential,
these bifurcations are not generic (which would imply period doubling)
but they are isochronous (i.e., each new-born orbit at the bifurcation
point has the same period as the parent orbit) \cite{nong}. (We have
tried to follow a sequence of period-doubling bifurcations in the HH
system. However, this soon leads to very long orbits that become
unstable very fast with increasing energy; their investigation after
more than three doublings has turned out to be numerically very
difficult. Also, the successive period-doubling bifurcations are not
all of pitchfork type but seem to alternate between pitchfork and
touch-and-go type.)

One important result of the Feigenbaum theory was to establish the
geometric progression of the bifurcation values of the system parameter
through a constant $\delta$ that turned out to be universal for a
certain class of maps. In the original work of Feigenbaum \cite{feig},
a dissipative one-dimensional map was investigated. Successive studies
in area preserving two-dimensional maps \cite{fei2,fei3} yielded a
different value of $\delta$. Translating to the present situation, we
take the energy $e$ as the system parameter and study the progression
of the energy intervals $1-e_n$ between the cumulation point $e=1$ and
the $n$-th bifurcation. Note, however, that the new orbits are born in
two different sets (see \fig{zoom} below): the stable rotations
R$_{2m-1}$ with two time orientations, and the unstable librations
L$_{2m}$ that come in degenerate pairs lying symmetrically to the the
saddle line containing the parent orbit ($m\geq3$). It is thus
necessary to study the corresponding sequences of bifurcations
separately, so that we have to determine the ratios
\be
\delta_n = \frac{1-e_n}{1-e_{n+2}}
\label{geome}
\ee
separately for odd and even $n$, and to see if the values of $\delta_n$
become constant for large $n$. Averaging our numerical values
$\delta_n$ from the range $7\leq n \leq 12$, we obtain $\delta_{o}=
37.623$ from the odd $n$ and $\delta_{e}=37.633$ from the even $n$. The
standard deviation from their mean value 37.628 is 0.082, so that the
difference between $\delta_{e}$ and $\delta_{o}$ is insignificant and
we may conclude that the mean value $\delta=37.628$ is unique. Below,
we shall determine analytically an asymptotic value for $\delta$ which
confirms this numerical result.

But let us first examine another important property of the Feigenbaum
scenario for area preserving two-dimensional maps: two independent
numerical constants $\alpha$ and $\beta$ were found to describe the
scaling of the fixed points at the bifurcations in the two directions
of the map \cite{fei3}. In our Hamiltonian system, Poincar\'e surfaces
of section -- chosen through the $u$ or the $v$ axis -- would provide 
us with two-dimensional area preserving maps and allow us to study the
evolution of the fixed points with energy. However, such a study is
again hampered by numerical inaccuracies. Instead, we found a
geometrical self-similarity of the periodic orbits born at the
bifurcations that reflects the fractal pattern of the fixed points in
the Poincar\'e maps and can be analyzed numerically with higher
accuracy.

\Figurebb{selfsim}{22}{279}{576}{577}{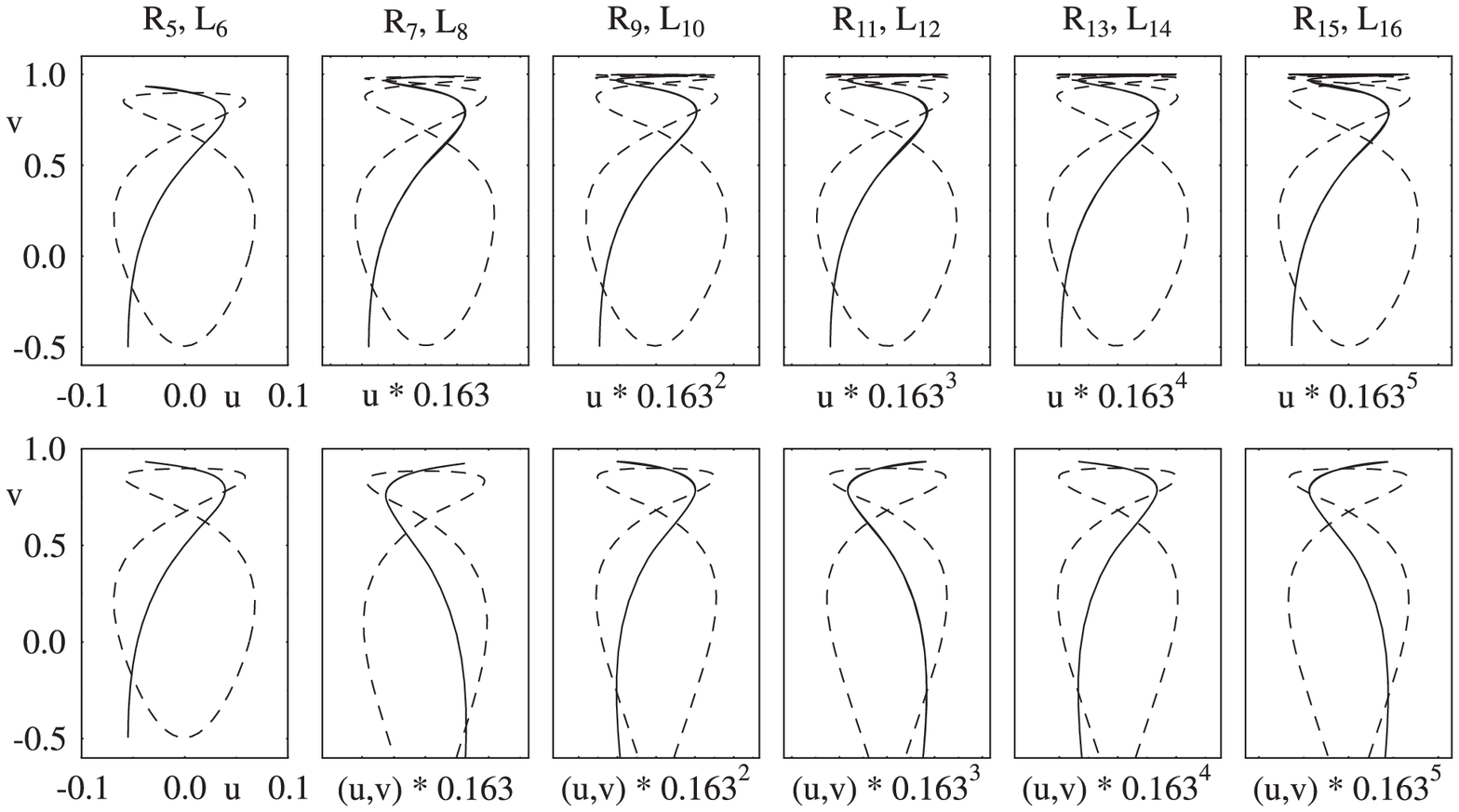}{8.4}{15.5}{
Orbits bifurcated from the vertical A orbit in the H\'enon-Heiles 
potential at energy $e=1$, shown with increasing Maslov indices from 
left to right. {\it Dashed lines:} rotations R$_{2m-1}$, {\it solid 
lines:} librations L$_{2m}$ ($m\geq3$). Only one libration orbit is 
shown for each L$_{2m}$; its partner is obtained by reflection at 
the vertical symmetry line $u=0$. {\it Top panels:} successive 
scaling of $u$ axis from left to right with the factor 0.163. {\it 
Bottom panels:} successive scaling of both axes with the same factor; 
along the $v$ axis only the top part starting from $v=1$ is shown.
}

In Fig.\ \ref{selfsim} we show the shapes of the orbits born at the
isochronous bifurcations of orbit A, with increasing Maslov indices
from left to right; all were evaluated at the barrier energy $e=1$. We
chose them here to be oriented along the $v$ axis on which their parent
orbit A is lying. The closer they are born to $e=1$, the smaller has
their amplitude in the transverse $u$ direction developed when they reach 
the barrier energy. Therefore, in the upper part of the figure, the $u$
axis has been zoomed by a factor 0.163 from each panel to the next, in
order to bring the shapes to the same scale. The orbits look
practically identical in the lower 97\% of their vertical range, but
near the barrier ($v=1$) they make one more oscillation in the $u$
direction in each generation. In the lower part of the figure, we have
zoomed also the $v$ axis by the same factor from one panel to the next
and plotted the top part of each orbit, starting from $v=1$. In these
blown-up scales, the tips of the orbits exhibit a perfect
self-similarity. The fact that the same scaling factor was used in both
directions means that we find the two scaling constants $\alpha$ and
$\beta$ to be identical here. We shall derive them analytically below
and show how they are related to $\delta$.

Note that although the parent orbit A becomes non-compact and
non-periodic for $e>1$, all periodic orbits bifurcated from it survive
up to arbitrary energy, becoming more and more unstable; in spite of
the non-compactness of the Hamiltonian \eq{hhxy} for $e>1$ they stay in
a finite region of space. At $e=1$, all the orbits R$_{2m-1}$ have
become inverse-hyperbolically unstable, whereas the L$_{2m}$ remain
direct-hyperbolically unstable. Vieira and Ozorio de Almeida
\cite{vioz} have also determined some of these orbits at $e>1$, both
numerically and semi-analytically using Moser's converging normal forms
near a harmonic saddle. They showed, like Davies \etal\ \cite{hhdb},
that with increasing energy $e>1$ these orbits come closer and closer
to the librating orbits $\tau$ that oscillate across the saddles. (In
Ref.\ \cite{hhdb}, our orbits R$_5$, L$_6$, R$_7$, and L$_8$ were named
$i_a$, $i_b$, $i_c$, and $i_d$, respectively, and $\tau$ was called S.)
In \fig{orbi11} we present the orbits R$_{11}$ and $\tau$ at
$e=1.00006625$, projected onto four different planes in the phase
space. The mixed projections $(u,p_u)$ and $(v,p_v)$ correspond exactly
to the results shown in Ref.\ \cite{vioz}. Note in the upper left panel
of the figure, how the R$_{11}$ orbit (solid line) winds around the
torus of the $\tau$ orbit (dotted line).

\Figurebb{orbi11}{35}{271}{564}{562}{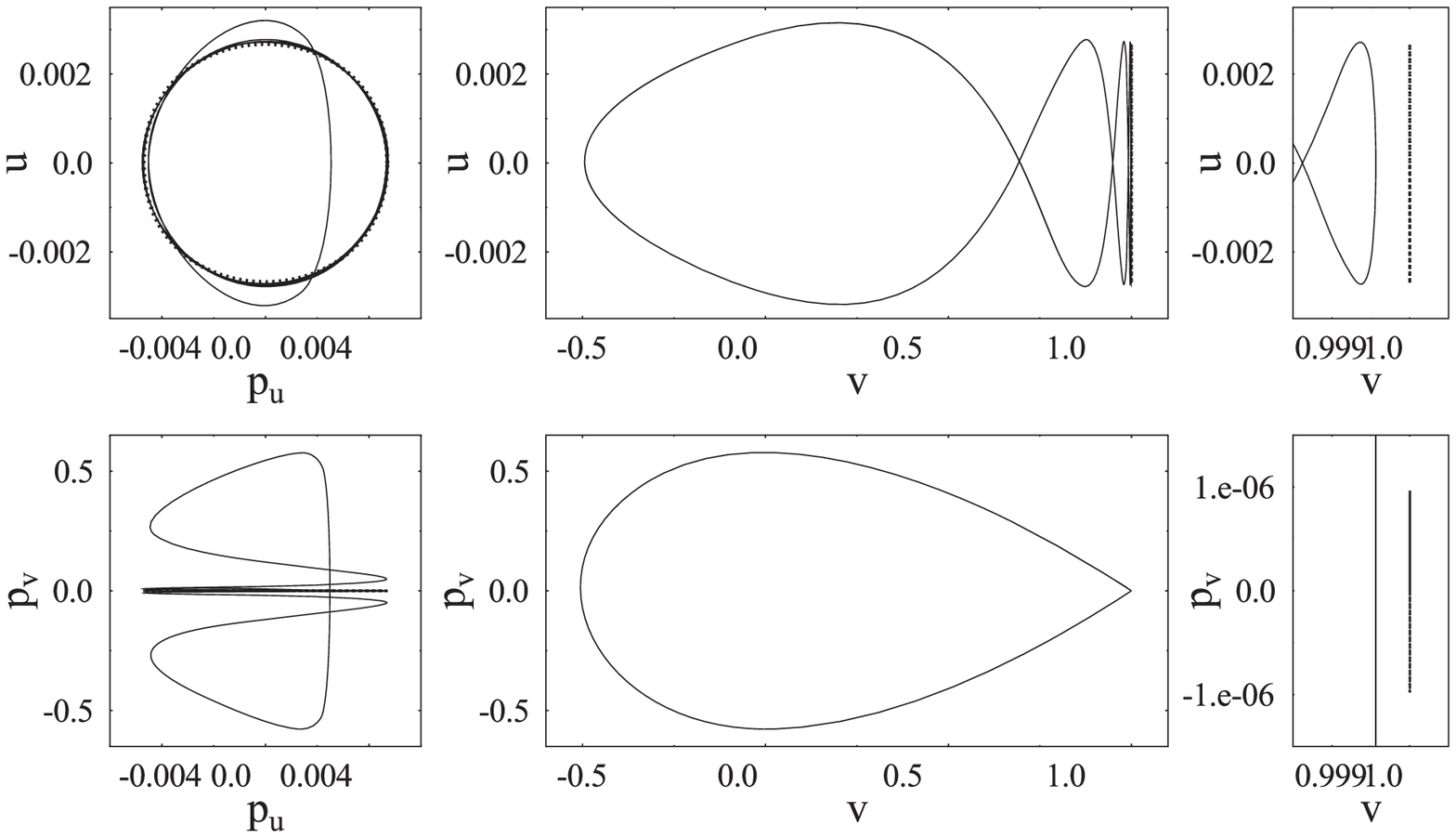}{8.5}{15}{
Periodic orbits evaluated at the energy $e=1.00006625$, projected onto
four different planes in the phase space. {\it Solid lines:} orbit
R$_{11}$, {\it dotted lines:} orbit $\tau$. Rightmost panels: zooming
the $v$ and $p_v$ axes near the saddle point.
}

We shall now proceed to derive the analytical values of the scaling
constants $\delta$, $\alpha$, and $\beta$. We give here only the main
idea of the derivation using intuitive arguments; a more detailed
mathematical analysis will be presented in a forthcoming publication
\cite{mbmm}. The key to the understanding of the geometric progression
of the bifurcation energies $e_n$ is a plot of tr\,M not versus energy
$e$ but versus the period $T$. This is shown in \fig{hhtrmt}. We see
that for large $T$, the quantity tr\,M of the orbit A exhibits a
perfectly periodic sinusoidal dependence on $T_A$ (which was noticed
already in Ref.\ \cite{hhdb}). The asymptotic period of these
oscillations is found here numerically to be $\Delta T = 3.6276\pm
0.0003$.

This result can be analytically derived by linearizing the equations of
motion \eq{hhxy} around the periodic orbit A which is a solution, e.g.,
with $u_A(t)=0$. The one-dimensional motion of this orbit in the $v$
direction is given by the scaled potential (cf.\ \fig{hhcont})
\be
6\,V_0(v) = 3\,v^2-2\,v^3.
\label{v0}
\ee

\Figurebb{hhtrmt}{55}{224}{559}{595}{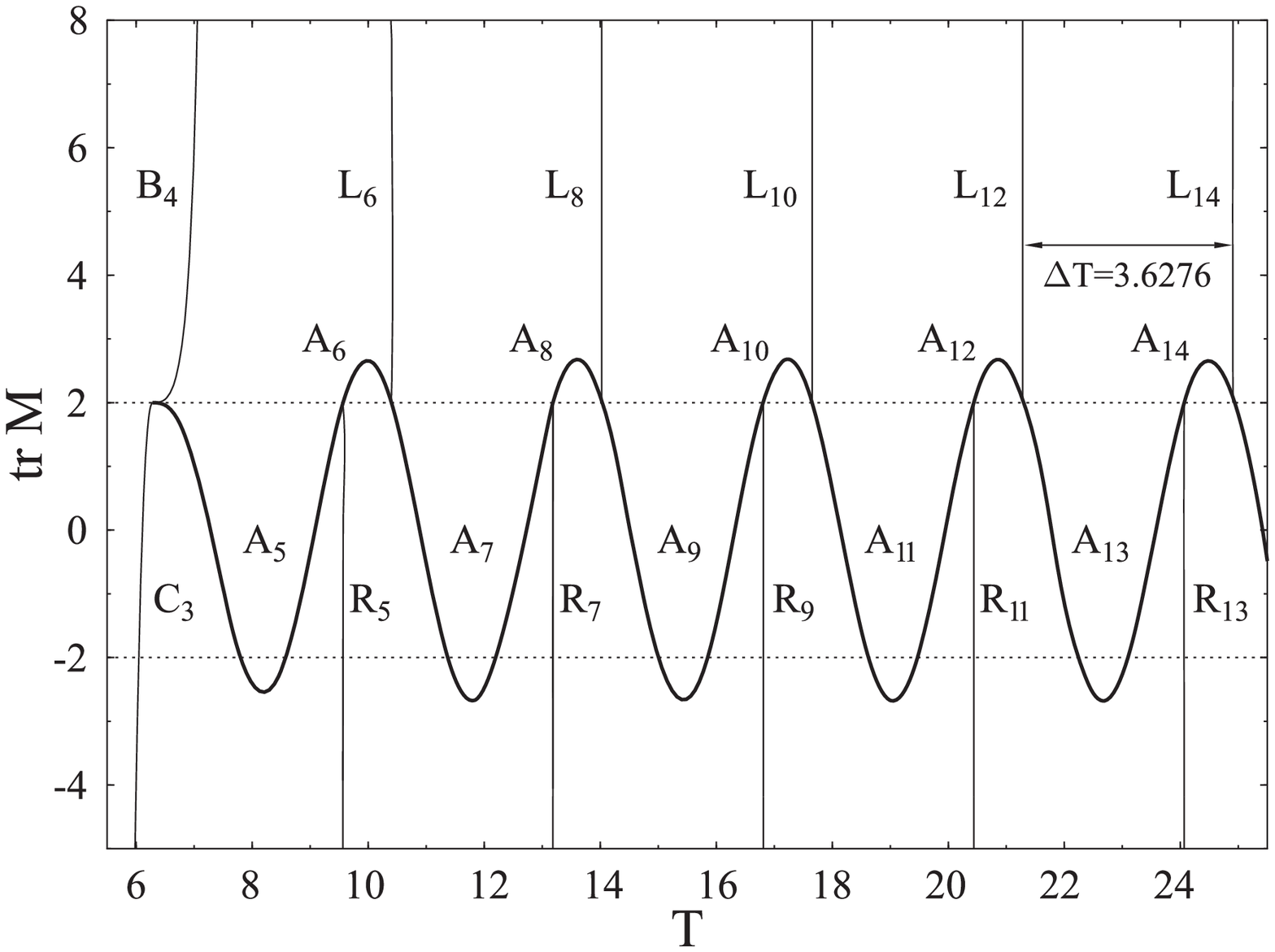}{8.5}{14.5}{
Trace of the stability matrix M of the orbits A (heavy line), B, C, and 
the orbits R$_{2m-1}$, L$_{2m}$ ($m\geq3$) born at successive pitchfork
bifurcations of orbit A in the H\'enon-Heiles potential, plotted versus 
their individual periods $T$. $\Delta T$ is the asymptotic period of 
the curve tr\,M$_A\,(T_A)$ for large $T_A$.
}

\noindent
Hence we find from energy conservation
\be
t\,(v) = \sqrt{3} \int_{v_1}^v \frac{ds}{\sqrt{e-3\,s^2+2\,s^3}}\,.
\ee
This integral can be expressed in terms of an elliptic integral of the
first kind, $F$, as
\be
t(v) = \sqrt{3/2}\int_{v_1}^v \frac{dv}{\sqrt{(v-v_1)(v_2-v)(v_3-v)}}
     = \sqrt{6/(v_3-v_1)}\,F(\gamma,k^2)\,.
\label{tofv}
\ee
Hereby $v_i$ ($i=1,2,3$) are the zeros of the equation $6\,V_0(v_i)=e$;
$v_1$ and $v_2$ are the turning points of the A orbit (see
\fig{hhcont}). The arguments of the elliptic integral are
\be
\gamma = \arcsin\sqrt{(v-v_1)/(v_2-v_1)}\,, \qquad
   k^2 = (v_2-v_1)/(v_3-v_1)\,.
\ee
The period of the A orbit thus becomes
\be
T_A=2\sqrt{3/2}\int_{v_1}^{v_2}\frac{dv}{\sqrt{(v-v_1)(v_2-v)(v_3-v)}}
   =2\sqrt{6/(v_3-v_1)}\,K(k^2)\,,
\label{ta}
\ee
where $K$ is the complete elliptic integral of the first kind which
diverges when the modulus $k^2$ becomes unity; this happens at $e=1$
where $v_2=v_3=1$. The function $t(v)$ in \eq{tofv} can be inverted in
terms of the Jacobi elliptic function $\rm{sn}(s,k^2)$ to yield the
exact solution for the $v$ motion of the A orbit:
\be
v_A(t) = v_1 + (v_2-v_1)\,\rm{sn}^2(s,k^2)\,,
\label{vaoft}
\ee
where $s$ is the scaled time variable
\be
s = t\sqrt{(v_3-v_1)/6}\,.
\ee

Linearizing Eq.\ \eq{hhxy} around the solution $u=0,\;v=v_A(t)$ yields
the equations of motion for small perturbations $\delta u(t)$, $\delta
v(t)$ around the A orbit:
\begin{eqnarray}
\delta \ddot u(t) + [1 + 2v_A(t)]\, \delta u(t) & = & 0\, ,
\label{linu} \\
\delta \ddot v(t) + [1 - 2v_A(t)]\, \delta v(t) & = & 0\, .
\label{linv}
\end{eqnarray}
These equations are of harmonic-oscillator type with time-periodic
frequencies, usually named after Hill \cite{hill} who investigated them
in connection with the lunar theory (cf.\ Ref.\ \cite{gubu}). With the
particular solution \eq{vaoft} for $v_A(t)$, Eq.\ \eq{linu} is actually
a special case of the Lam\'e equation (see Ref.\ \cite{mawi} for an
extensive discussion of Hill's and related equations). It is from this
equation that the stability of the A orbit towards small perturbations
$\delta u$ can be derived. Magnus and Winkler \cite{mawi} have given an
iterative scheme to solve Hill's equation using the Fourier expansion
of the time-periodic coefficient, which has been used to calculate the
stability of a linear periodic orbit in the diamagnetic Kepler problem
\cite{edmo,maod}. We shall present the application of this procedure to
the present Hamiltonian elsewhere \cite{mbmm}, and just anticipate here
that the result for tr\,M$_A$ is a series
\be
{\rm tr\,M}_A\,(T_A) = 2\,\cos\,(\omega_\perp T_A) + \dots\,,
\label{trmta}
\ee
where the dots indicate correction terms coming from the higher Fourier
components of $v_A(t)$ in \eq{vaoft}. All corrections have the same
period $2\pi/\omega_\perp$ as the leading term in \eq{trmta} and change
therefore only the amplitude and the phase of the oscillations in 
tr\,M$_A$. The
frequency $\omega_\perp$ is given by the constant term of the Fourier
expansion, which equals the time average of the coefficient in
\eq{linu} over the period $T_A$; in the limit $e\rightarrow 1$ it goes
to a constant:
\be
\omega_\perp^2 = \langle\, [1 + 2\,v_A(t)]\; \rangle_{T_A} \;
           \longrightarrow \; 3\,.
\ee
From this, we immediately obtain the asymptotic period
\be
\Delta T=
2\pi/\omega_\perp = 2\pi/\!\sqrt{3} = 3.6275987\dots
\label{delt}
\ee
which is found approximately from the numerical curve tr\,M$_A\,(T_A)$
shown in \fig{hhtrmt}.

This result can be intuitively obtained by the following reasoning.
Note that the orbit A spends most of its time near the saddle point
where $v=1$; this is the more true the closer the energy comes to
$e=1$. Replacing $v_A(t)$ by its saddle-point value, the coefficient in
\eq{linu} becomes $\omega_\perp^2 = 1+2\,v_A = 3$. We are, in fact,
just speaking in this lowest-order approximation of a harmonic
oscillation transverse to the A orbit
\be
\delta u(t) = u_0 \cos\,(\omega_\perp t+\phi)
            = u_0 \cos\,(\!\sqrt{3}\,t+\phi)\,,
\label{deluoft}
\ee
with a constant frequency $\omega_\perp$ given by the curvature of the
HH potential at the saddle:
\be
\omega_\perp^2=\left.\partial^{\,2}V(u,v)/\partial u^2\right|_{u=0,v=1}
              =3\,.
\label{omega}
\ee

\Figurebb{orbi10}{90}{60}{795}{550}{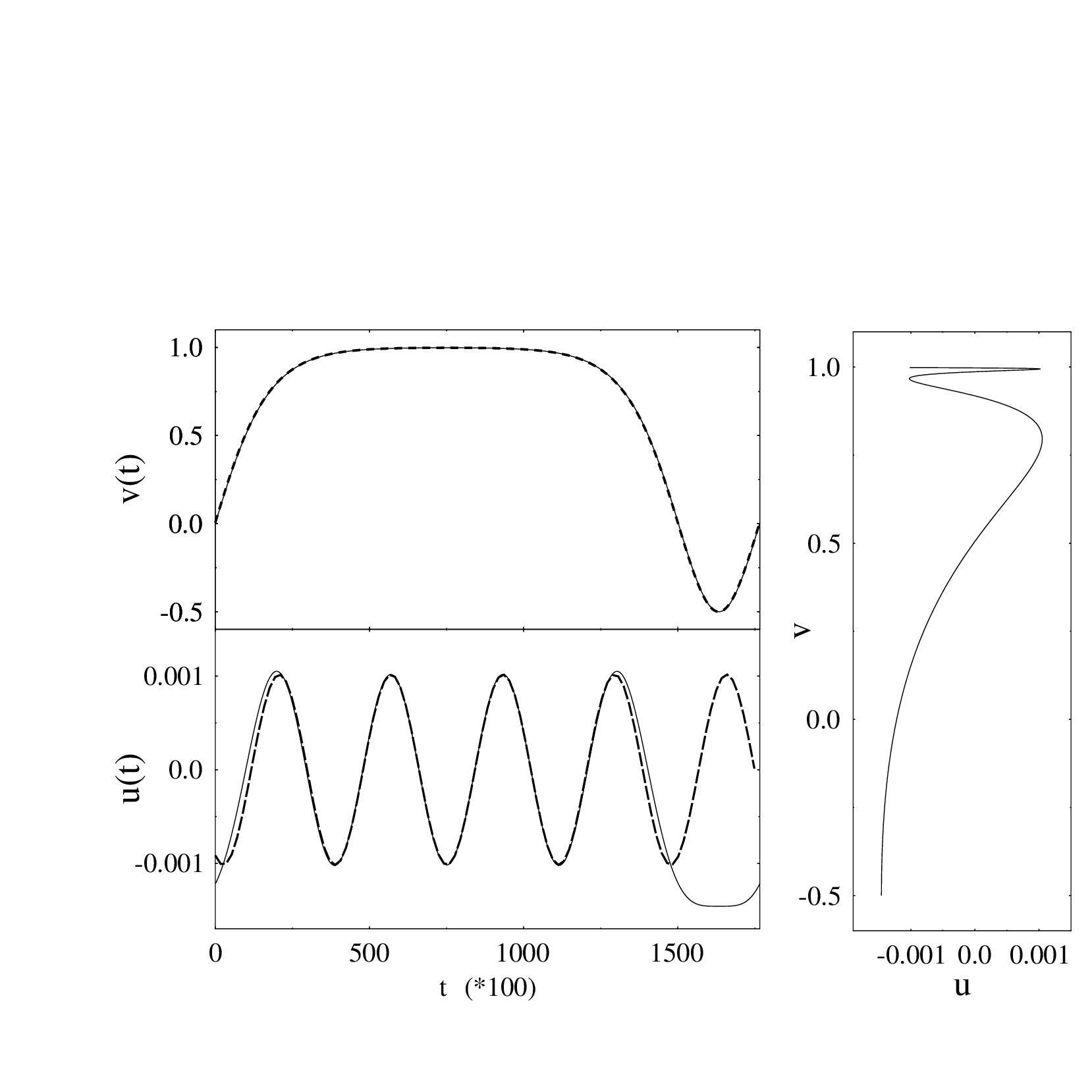}{8.5}{15.0}{
Orbit L$_{10}$ evaluated at $e=1$. {\it Right panel:} shape of the
orbit in the $(u,v)$ plane. {\it Left panels:} numerical results for
$v(t)$ (above) and $u(t)$ (below) versus time $t$ (in steps of 0.01),
shown by solid lines. Dashed line above: $v_A(t)$ of parent orbit A
taken at the bifurcation energy $e_{10}$. Dashed line below:
approximation \eq{deluoft} with $u_0=0.001017$ and the phase $\phi$
fitted to the numerical result.
}

\noindent
In this limit, the stability matrix of the A orbit has exactly the
value tr\,M$_A\,(T_A) =2\cos\,(\omega_\perp T_A)$ (see, e.g., Ref.\
\cite{book}, Appendix C.2.1), in agreement with the leading term of the
expansion \eq{trmta}. That $\delta u(t)$ is harmonic with frequency
$\omega_\perp=\sqrt{3}$ over most of the time period $T_A$ is clearly
seen in the results presented in \fig{orbi10}, where we show the
numerical solutions for $u(t)$ and $v(t)$ of the orbit L$_{10}$
evaluated at the saddle energy $e=1$. The transverse $u$ motion is,
indeed, fitted extremely well over most of the period by Eq.\
\eq{deluoft} shown by a dashed line; the value of $u_0$ will be
determined below.

The periods $T_n$ at which the bifurcations occur are thus given
asymptotically by $T_{2m-1} \sim a + m\,\Delta T$ and  $T_{2m} \sim b +
m\,\Delta T$ (see \fig{hhtrmt}), with $\Delta T$ given by Eq.\
\eq{delt} and some constants $a$, $b$. The bifurcation energies $e_n$
are now easily found by the asymptotic expansion of the period $T_A$
\eq{ta} near $e=1$, i.e., near $k^2=1$, where it diverges like
\be
T_A \;\sim\; \ln\,(4/\!\sqrt{1-k^2})+\dots
    \;\sim\; \ln\,(432/\epsilon)+\dots\,,
\ee
with $\epsilon=1-e$. (This is easily derived from a Taylor expansion of
the turning points $v_i$ and the quantity $k^2$ in powers of
$\epsilon$.) Hence we find, for both even and odd $n$,
\be
1-e_n \;\sim\; 432\,e^{-T_n} \;\propto\; e^{-n\Delta T/2}
      \;\propto\; e^{-n\pi/\omega_\perp}\,.
\label{inten}
\ee
We thus obtain with \eq{geome} the analytical value of the asymptotic
energy scaling constant $\delta$
\be
\delta = e^{2\pi/\omega_\perp} = e^{2\pi/\!\sqrt{3}} = 37.622367\dots
\label{delta}
\ee
which confirms the approximate numerical result given after Eq.\
\eq{geome}.

\newpage

The spatial scaling constants $\alpha$ and $\beta$ describing the
self-similarity of the new periodic orbits are derived along the same
lines. In the limit $v_A(t)=1$, the equation for $u(t)$ is harmonic
even in the full equations of motion \eq{hheom} and decouples from that
for $v(t)$. Furthermore, the equation for $v(t)$ is of second order in
$u$ so that to lowest order in small $u$ oscillations, $v(t)$ is not
changed at all. Consequently, as long as the amplitude $u_0$ of the
transverse motion $u(t)$ of the bifurcated orbits remains small, their
$v$ motion is ``frozen'' at the bifurcation point and given by the
solution $v_A(t)$ in \eq{vaoft}, evaluated at $e=e_n$. The transverse
motion $u(t)$ of the new orbit therefore carries off all the extra
energy $e-e_n$ available above the bifurcation. That the $v$ motion of
these orbits is frozen above the bifurcation energies is seen in
\fig{hhtrmt} by the fact that they appear as almost vertical lines (for
large enough $n$), which means that their periods are practically
constant. Also, in \fig{orbi10} we see that $v(t)$ of the orbit
L$_{10}$ obtained numerically at $e=1$ (solid line) is, indeed,
identical to $v_A(t)$ evaluated at $e_{10}$ (dashed line, hardly 
distinguishable from the solid line). As a consequence of the frozen 
$v$ motion, the energy of the $u$ motion available for the new orbit at 
the saddle-point energy is just $1-e_n$. In the harmonic approximation 
\eq{deluoft}, this energy is equal to
\be
1 - e_n = 3\,({\dot u}^2+\omega_\perp^2u^2) = 3\,\omega_\perp^2 u_0^2
        = 9\, u_0^2\,.
\ee
Hence we find that the transverse amplitude of the new orbit at the
saddle is given by
\be
u_0 = \sqrt{1-e_n}/3\,.
\label{u0}
\ee
Using the numerical result $(1-e_{10})= 9.305\times10^{-6}$, the
amplitude for the orbit L$_{10}$ becomes from the above relation
$u_0=0.001017$ which is, indeed, the value that fits the numerical
result for $u(t)$ in \fig{orbi10}. Now, from Eq.\ \eq{u0} we find that
the ratio of the amplitudes $u_0$ of two successive generations of
orbits for $n\rightarrow\infty$ tends to
\be
\sqrt{(1-e_n)/(1-e_{n+2})} \; \longrightarrow \; \alpha =
\sqrt{\delta} = e^{\pi/\!\sqrt{3}}.
\label{alpha}
\ee
This confirms the numerical scaling constant $1/\alpha=0.1630335\dots$
approximately used for the $u$ scaling in \fig{selfsim}. The scaling
constant $\beta$ for the $v$ direction, finally, is obtained from
expanding the potential \eq{v0} around the barrier at $v=1$:
\be
6\,V_0(v) = 1 - 3\,(v-1)^2 + \dots
\label{v0app}
\ee
Since the $v$ motion of the $n$th orbit is frozen, its tip near the
barrier is just the turning point $v_2$ evaluated at $e_n$. Equating
the frozen energy $e_n$ with $6\,V_0(v_2)$ to leading order in 
\eq{v0app}, we find the distance $1-v_2$ of the tip from the saddle 
point to be
\be
1-v_2 = \sqrt{(1-e_n)/3}\,.
\ee
Its ratio for two successive generations of orbits thus goes with
$n\rightarrow\infty$ to the same limit as that found in \eq{alpha} for
the $u$ scaling. Hence $\beta=\alpha$, as found numerically in
\fig{selfsim}.

This concludes the derivation of our main result: the analytical value
\eq{delta} of $\delta$, and the relation $\alpha=\beta=\sqrt{\delta}$.
There are further interesting observations, to be interpreted in future
work \cite{mbmm}. For large $n$ the actions of the bifurcated orbits at 
the saddle
energy $e=1$ all tend to the action of the orbit A which is known
analytically \cite{hh1}: $S_A=6/5$. Also at $e=1$, the values for tr\,M
of all the orbits R$_{2m-1}$ tend to the same value $-4.183$ (see their
intersecting lines in \fig{zoom}), whereas those of the L$_{2m}$
intersect at the value $+8.183$.

Finally, we add a remark about the $\tau$ orbits that exist only for
$e>1$. They are oscillations transverse to the saddles (see
\fig{orbi11}, in particular the upper right panel). In the limit
$e\rightarrow +1$ they become one-dimensional harmonic oscillations
with the frequency $\omega_\perp=\sqrt{3}$. Their period is, in this
limit, equal to $T_\tau=2\pi/\omega_\perp$ which is identical to
$\Delta T$ in \eq{delt}. Their transverse motion feels a negative
curvature (corresponding to the passage over the saddle parallel to the
orbit A) with the value
\be
- \omega_\parallel^2 = \partial^{\,2} V(u,v)/\partial v^2|_{u=0,v=1}
                     = - 1\,,
\label{ompar}
\ee
as is seen directly from Eq.\ \eq{linv} with $v_A=1$. They are
therefore unstable, and their stability matrix has the trace (cf.\
Ref.\ \cite{book}, Sect.\ 5.6.3)
\be
{\rm tr}\,M_\tau = 2\,\Cosh\,(\omega_\parallel T_\tau)
                 = 2\,\Cosh\,(2\pi/\!\sqrt{3})\,.
\ee
One of the eigenvalues of M$_\tau$ is thus
$\lambda=e^{2\pi/\!\sqrt{3}}$ with the Lyapounov exponent
$\chi=2\pi/\!\sqrt{3}$ (or $\sigma=\chi/T_\tau =\omega_\parallel=1$).
$\lambda$ is here identical with the scaling constant $\delta$
\eq{delta} for the bifurcation energies $e_n<1$ at which the orbits
R$_{2m-1}$ and L$_{2m}$ are born. This exhibits once more the intimate
connection \cite{hhdb,vioz} between the two types of periodic orbits 
near the threshold $e=1$, already displayed in \fig{orbi11}.

\bs

{\bf 3. Results for other two-dimensional potentials}

\ms

\noindent
We have investigated various two-dimensional potentials with one
or more harmonic saddles and orbits that oscillate along
straight lines towards the saddles. In all cases, we could find
the same type of bifurcation cascades and the same self-similarity
of the new-born orbits obeying the relation $\alpha=\beta=\sqrt{\delta}$.
We shall give three examples below.

\bs

{\bf 3.1. A simple integrable case}

\ms

\noindent
A separable potential with a minimum and one saddle is obtained if one
omits the term proportional to $x^2y$ of the HH potential \eq{hhxy} to
get
\be
V(x,y) = \frac12\,(x^2+y^2)-\frac13\,\varepsilon\,y^3\,,
\label{integ}
\ee
and otherwise proceeds exactly in the same way as in Sect.\ 2. The
straight-line A orbit along the $v$ axis here sees the same potential
as the A orbit in the HH potential and thus has the same solution
$v_A(t)$ given in Eq.\ \eq{vaoft}. Since the potential separates in $u$
and $v$, the system is integrable. Nevertheless, the A orbit undergoes
an infinite cascade of bifurcations cumulating at the energy $e=1$. The
$u$ motion is strictly harmonic with period $T_0=2\pi$, corresponding
to $\omega_\perp=1$, and the trace of the stability matrix of A is
exactly tr\,M$_A\,(T_A)=2\,\cos\,(T_A)$. The bifurcations thus occur at
the periods $T_n=2n\pi$ with $n=1$, 2, 3, $\dots$ This leads to a
scaling constant $\delta=e^{2\pi}$ for the progression of the
bifurcation energies $e_n$. The new periodic orbits born at the
bifurcations are here degenerate families with tr\,M $=2$ whose tips
scale in $u$ and $v$ with the constant $\sqrt{\delta}=e^\pi$.

\newpage
%\bs

{\bf 3.2. The Barbanis potential}

\ms

\noindent
Omitting the term proportional to $y^3$ in the HH potential yields a
potential that has been studied in 1966 by Barbanis \cite{w1}:
\begin{equation}
V(x,y) = \frac12\,(x^2+y^2) - \varepsilon \, yx^2 \,.
\label{w1xy}
\end{equation}
This potential has a minimum at $x=y=0$ and two saddles at the energy
$E^*=1/8\varepsilon^2$. There are two straight-line periodic orbits A
oscillating through the minimum towards the saddles. Scaling the
coordinates with a factor $\varepsilon$ and rotating the coordinate
system such that one of the saddle lines becomes the horizontal $u$
axis, the potential is
\begin{equation}
 V(u,v) = \frac{1}{2}\,(u^2+v^2) - \frac{1}{3\sqrt{3}}
          \left( 2\, u^3 - 3\, u v^2 + \sqrt{2}\, v^3 \right).
\label{scalv}
\end{equation}

\Figurebb{w1orbs}{30}{137}{577}{690}{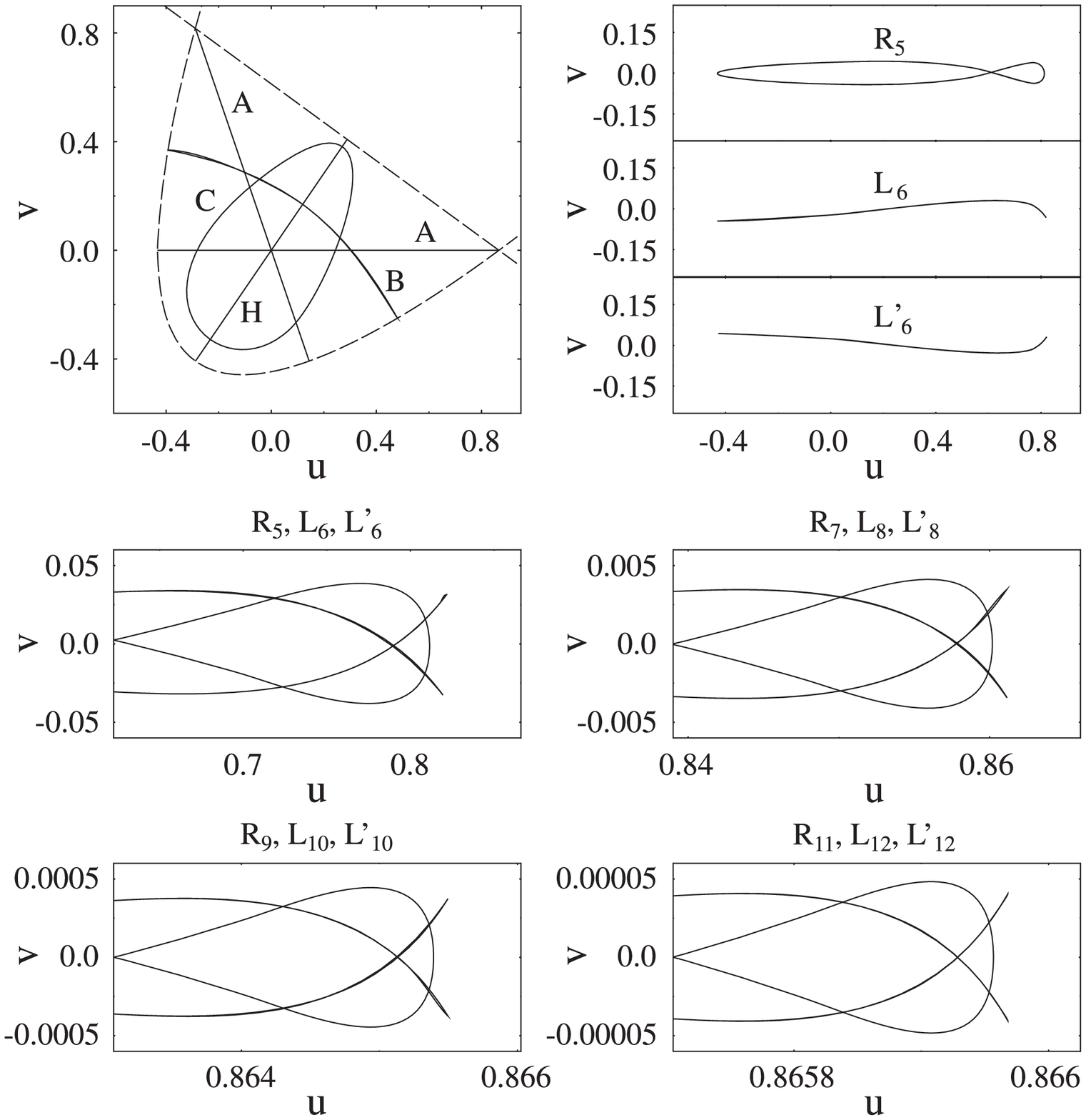}{13}{15.0}{
Periodic orbits in the potential of Barbanis \cite{w1}, evaluated at
the saddle-point energy $e=1$. {\it Top panel left:} primitive orbits A
(in two positions), B, C, and H in the ($u,v$) plane, shown by solid
lines. The dashed lines give the equipotential curves at $e=1$. {\it
Top panel right:} the first generation of orbits bifurcated from the
horizontal A orbit, shown on the same scale. {\it Lower panels:}
extreme right tips of four generations of bifurcated orbits. The
position of the saddle is at $u=\sqrt{3}/2 = 0.866025$ which
corresponds to the right margins. The scaling factor between successive
generations is 0.1084 in both directions.
}

\fig{w1orbs} shows the primitive periodic orbits in this potential. In
the top left panel, we see the equipotential curves for $e=E/E^*=1$ by
dashed lines: they form a parabola and a straight line intersecting at
the two saddles. The solid lines indicate the shortest periodic orbits
evaluated here at $e=1$. Note that the two straight saddle lines that
contain the A orbits are no symmetry axes (in contrast to the HH
potential). The potential \eq{scalv} has only one symmetry line which
halves the angle between the saddle lines and contains a straight-line
librating orbit H$_4$ that is unstable at all energies. There is one
curved librating orbit B$_4$ that oscillates transverse to the symmetry
axis and is also unstable at all energies (like the orbits B in the HH
potential). C is a rotating orbit that remains stable up to $e=2.415$.
The orbits A have the same behavior as in the HH potential and create
an infinite set of new orbits at isochronous pitchfork bifurcations.

Since the potential is not symmetric about the saddle lines, the new
libration orbits born at the bifurcations come in pairs with equal
Maslov indices $2n$ but shapes, actions and stabilities that differ
at energies $e>e_{2n}$; we call them here L$_{2n}$ and L'$_{2n}$. The
new rotations R$_{2n-1}$ again have two time orientations and thus a
discrete degeneracy of two. The shapes of these new orbits resemble
much those in the HH potential (apart from the asymmetry of the
unstable pairs). Their self-similarity is shown in the lower four
panels of \fig{w1orbs}, where their tips close to the saddle point
$u=\sqrt{3}/2$ are displayed for each generation. The panels from each
generation to the next are scaled by a factor 0.1084 both in $u$ and
$v$ direction. The bifurcation energies $e_n$ are again found to
cumulate in a geometric progression, with a numerical scaling constant
$\delta = 85.1 \pm 0.5$. 

The values of tr\,M of all the primitive 
orbits are shown in \fig{w1trmt} as functions of their period $T$. We 
find again a periodic behavior of tr\,M$_A\,(T_A)$, here with period 
$\Delta T=4.443$. 

\Figurebb{w1trmt}{46}{216}{567}{606}{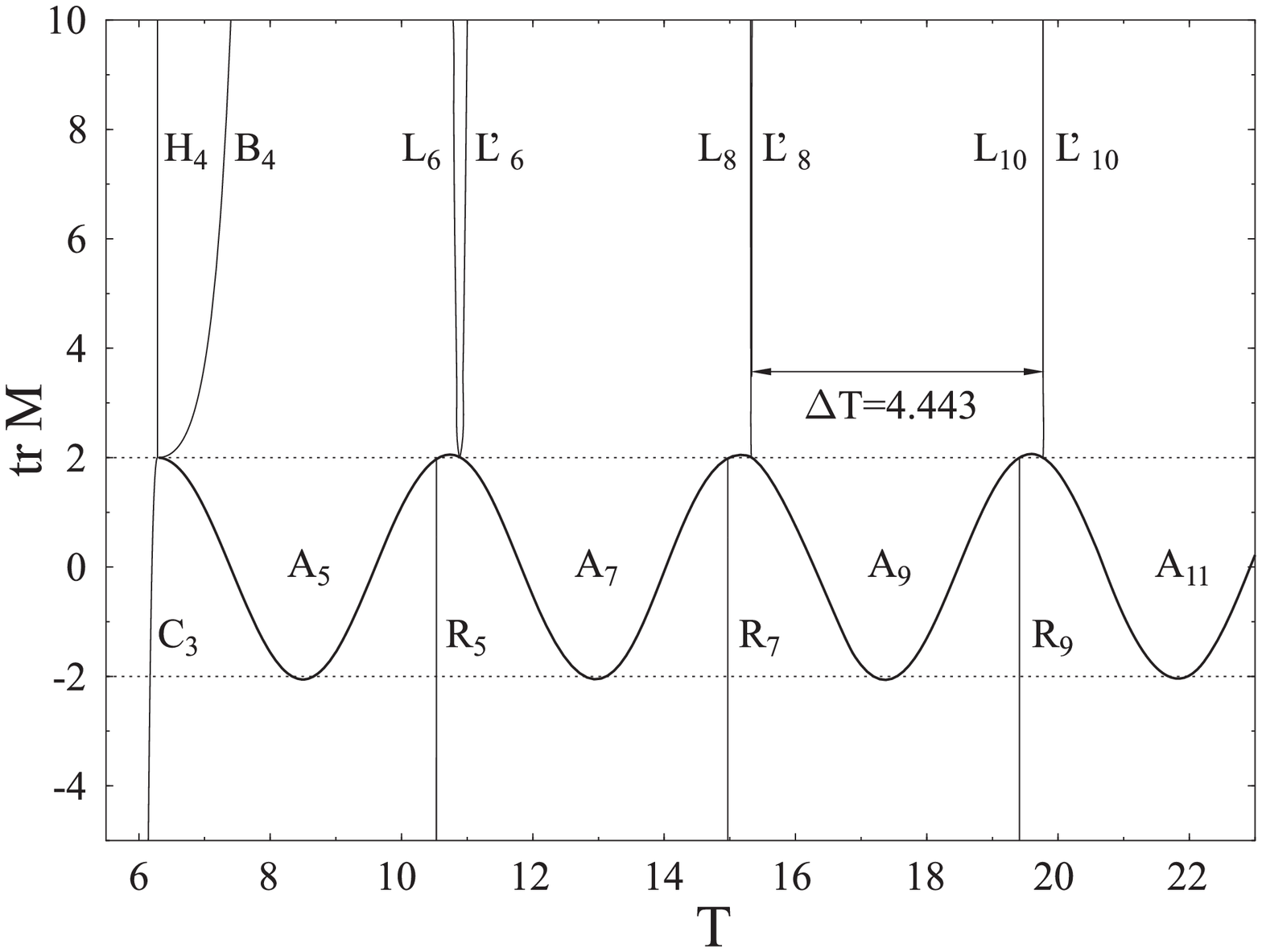}{8.2}{15.0}{
Trace of stability matrix M of orbits A (heavy line), B, C, H, and the
new orbits born at successive pitchfork bifurcations of orbit A in the 
Barbanis potential, plotted versus their individual periods $T$.
}

\noindent
The analytical values of these constants are found exactly in the same 
way as for the HH potential. In fact, the one-dimensional barrier seen 
by the A orbits in the Barbanis potential is, apart from a trivial 
scaling factor $\sqrt{3}/2$, identical to that in the HH potential. The
frequency of the transverse oscillations across the saddles here is
found to be $\omega_\perp=\sqrt{2}$, so that $\Delta T = \sqrt{2}\pi =
4.4428829\dots$\,, and the scaling constant $\delta$ of this potential
becomes
\be
\delta = e^{\sqrt{2}\pi} = 85.019695\dots
\ee
The scaling constants for the self-similar tips of the periodic orbits
are therefore
\be
\alpha = \beta = \sqrt{\delta} = 1/0.10845266\dots
\ee
Both these values confirm the numerically determined constants.

\bs

{\bf 3.3. The quartic H\'enon-Heiles potential}

\ms

\noindent
In Refs.\ \cite{hhun,pert} a quartic H\'enon-Heiles potential was
investigated that has a four-fold discrete rotational symmetry and four
saddle points. It is given, in scaled coordinates, by
\be
V(u,v) = \frac12\,(u^2+v^2) - \frac14\,(u^4+v^4) + \frac32\,u^2v^2\,.
\label{r4xy}
\ee
We refer to the above literature for a discussion of the shortest
periodic orbits (which we have renamed here to simplify the notation).
We show the equipotential curves and the orbits A, B, and C in the top
left panel of \fig{r4orbs}. The orbits A have the same behavior as
those in the standard HH potential, except that they oscillate between
two opposite saddle points. They undergo again an infinite series of
isochronous pitchfork bifurcations. The difference to the standard HH
case is that here the nature of the new orbits born at the bifurcations
alternates between rotations R$_n$ and librations L$_{n'}$ also amongst
the stable orbits (odd $n$) and the unstable orbits (even $n$), as is
shown for the orbits R$_5$, L$_6$, L$_7$, and R$_8$ labeled explicitly
in \fig{r4orbs}. As a consequence, the self-similarity of their tips
near the saddles becomes apparent only over two generations.
Correspondingly, each of the lower four panels contains the tips of two
successive generations of orbits, and the scaling from one panel to the
next is done with the factor $\alpha^2=\beta^2=\delta$. The frequency
of the small oscillations across the saddles here is $\omega_\perp=2$,
and the period of tr\,M$_A\,(T_A)$ becomes asymptotically $\Delta T =
2\pi/\omega_\perp=\pi$. The expansion of $T_A$ near the saddle energy
gives here, with $\epsilon=1-e$,
\be
       T_A \;\sim\; \sqrt{2}\,\ln\,(64/\epsilon) + \dots
\label{tar4}
\ee
For the bifurcation energies $e_n$ we have thus asymptotically (note
the extra factor $\sqrt{2}$)
\be
1-e_n \;\sim\; 64\, e^{-T_n/\!\sqrt{2}}.
\ee
With $T_n \sim n\pi/\omega_\perp=n\pi/2$, we find for the energy 
scaling constant
\be
\delta = e^{\pi/\!\sqrt{2}} = 9.2206125\dots
\label{r4del}
\ee
Its inverse is $1/\delta = 1/\alpha^2 = 0.108452664\dots$\,, i.e., the
scaling factor used in \fig{r4orbs}.

\Figurebb{r4orbs}{70}{116}{571}{730}{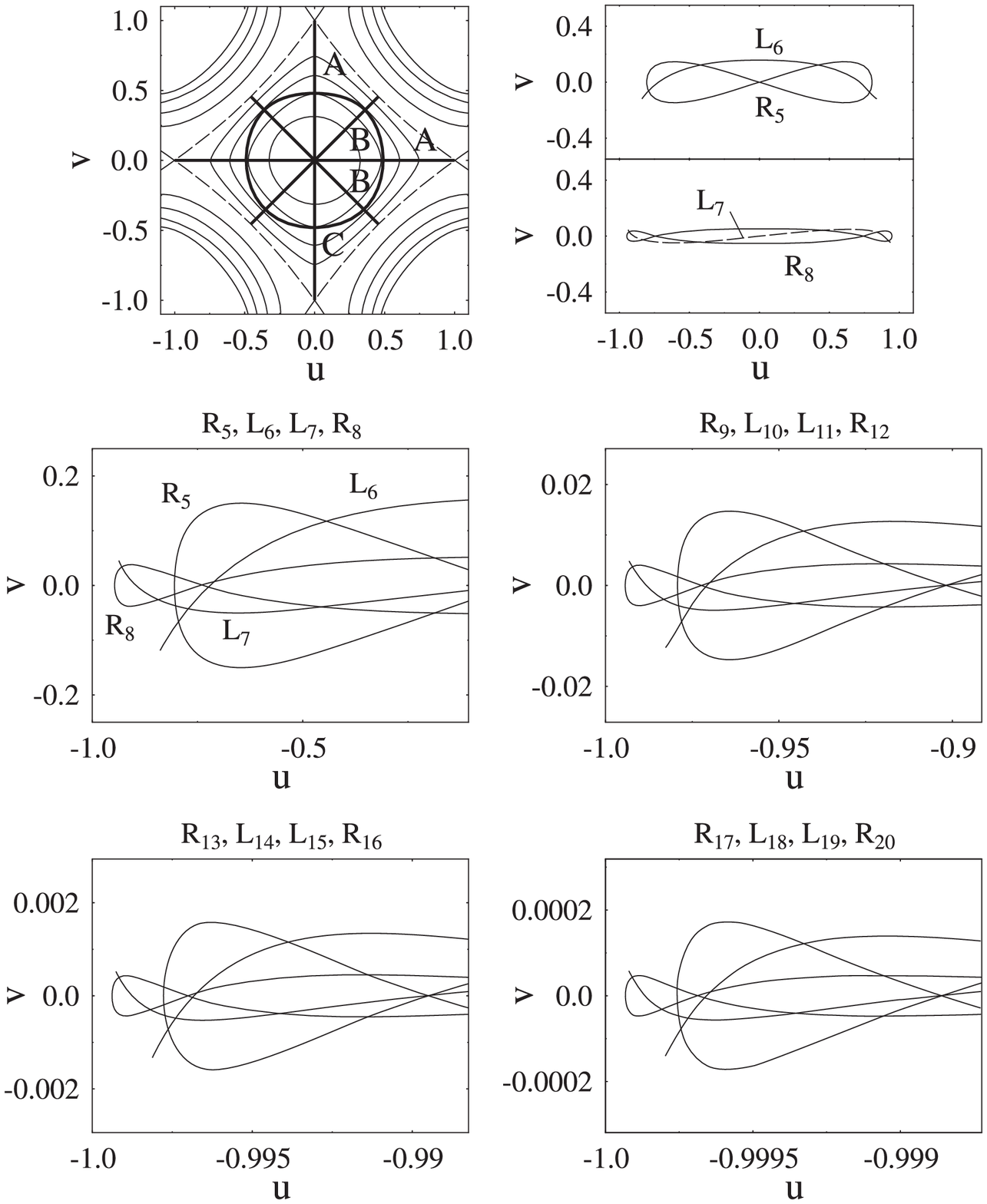}{14.8}{15.0}{
Periodic orbits in the quartic H\'enon-Heiles (HH4) potential,
evaluated at $e=1$. {\it Upper left panel:} equipotential curves (thin
lines; dashed for $e=1$) and the primitive orbits A, B, and C (heavy
lines). {\it Upper right panel:} the first two generations of
bifurcated horizontal orbits R$_5$, L$_6$ (above) and L$_7$, R$_8$
(below; orbit L$_7$ is dashed), drawn on the same vertical and
horizontal scales as the orbits in the upper left panel. {\it Lower
four panels:} Tips of the bifurcated orbits R$_5$ - R$_{20}$ near the
left saddle ($u=-1)$, plotted on increasingly zoomed scales. The
scaling factor from each pair of generations to the next is 0.10845 in
both directions.
}

The numerical iteration of the bifurcation energies $e_n$ was easier
in this potential. The average value of the $\delta_n$ given by Eq.\
\eq{geome} in the range $7\leq n \leq 16$ gives a mean value of $\delta
= 9.2203$ with a standard deviation of 0.02. The asymptotic value of
$\Delta T=\pi$ is reached for the interval $T_{19}-T_{17}$ with an
accuracy of 7 digits.

\newpage
%\bs

{\bf 4. Summary and conclusions}

\ms

\noindent
We have investigated cascades of isochronous pitchfork bifurcations of
straight-line lib\-rational orbits in the H\'enon-Heiles and similar
two-dimensional potentials possessing one or more harmonic saddles.
The bifurcation energies cumulate at the saddle-point energy (scaled
to be $e=1$); their geometric progression yields a scaling constant
$\delta$ that can be calculated analytically in terms of the potential
parameters. The periodic orbits born at the successive bifurcations
exhibit a self-similarity corresponding to scaling constants $\alpha$
and $\beta$ in the two spatial directions that turn out to be identical
and equal to $\alpha=\beta=\sqrt{\delta}$.

This result applies to a whole class of Hamiltonian systems with
harmonic saddles and straight-line librational orbits oscillating
towards the saddle points. The trace of the stability matrix of these
orbits as a funcion of their period $T_A$, tr\,M\,($T_A$), oscillates
with an asymptotic periodicity $\Delta T=2\pi/\omega_\perp$, where
$\omega_\perp^2$ is the transverse curvature at the saddles, see Eq.\
\eq{omega}. Combining this with the asymptotic energy dependence $T_A
\sim d\ln[c/(1-e)]$, one finds the asymptotic values $e_n$ of the
bifurcation energies. From those, the analytic form of the scaling
constant $\delta$ in terms of the parameters $\omega_\perp$ and $d$ is
found to be $\delta= e^{2\pi/\omega_\perp d}$. The Lyapounov exponent
of the new unstable orbits $\tau$ librating across the saddles for
$e>1$ is, in the limit $e\rightarrow +1$, given by $\chi=2\pi
\omega_\parallel/\omega_\perp$ (or $\sigma=\chi/T_\tau=
\omega_\parallel$), where $-\omega_\parallel^2$ is the negative
parallel curvature at the saddles, see Eq.\ \eq{ompar}.

The well-known Feigenbaum scenario in two-dimensional area preserving
maps \cite{fei2,fei3} differs from that studied here in three respects.
First, the bifurcations discussed there are successive period-doublings
which form a fractal tree. Second, their constants $\delta$, $\alpha$,
and $\beta$ are all different from each other and only known
numerically. Third, these constants appear to be universal for a whole
class of quadratic maps, whereas in the present case the constant
$\delta$ depends explicitly on the parameters of the potential. We
should bear in mind that the Poincar\'e maps corresponding to the
Hamiltonian systems studied here are, of course, no simple quadratic
maps.

An isochronous pitchfork bifurcation cascade of a linear periodic orbit 
has been found also in the diamagnetic Kepler problem \cite{maod,main}. 
This orbit does not approach a saddle point but becomes infinitely long 
in the limit $E\rightarrow -\,0$ where its bifurcations cumulate. The 
self-similarity of the periodic orbits born at the bifurcations has not 
been investigated; the progression of the bifurcation energies is 
easily found to yield the constant $\delta=1$ (cf.\ Refs.\ 
\cite{sum,mai2}). The same value $\delta=1$ is also found for the 
bifurcations of the short diameter orbit in the ellipse billiard, which 
are $m$-uplings with $m\geq2$ cumulating in the limit $m\rightarrow
\infty$ at zero excentricity (cf.\ Ref.\ \cite{elli} and Ref.\ 
\cite{book}, problem 5.3). 

It will be interesting to study also the modifications arising in 
connection with non-harmonic saddles, or with curved librating orbits 
approaching a saddle.

\bs

{\bf Acknowledgments}

\ms

\noindent
I am very grateful to Mitaxi Mehta for illuminating discussions and a 
critical reading of the manuscript, and to Kaori Tanaka for considerable
improvements of our numerical POT code in an ongoing collaboration 
\cite{kabr}. I also acknowledge encouraging discussions with S. Fedotkin, 
J. Kaidel, A. Magner, M. V. N. Murthy, J. M. Rost, and M. Sieber.

\bs
\bs

{\bf Note added in proof:}

\bigskip

\noindent
The transverse motion $u(t)$ [or $v(t)$ in Sects.\ 3.2 and 3.3] of the 
bifurcated orbits R$_n$, L$_n$ is given, close enough to their bifurcation 
energies so that its amplitude remains small, by the periodic Lam\'e 
functions$^1$ Ec$_p^m(at)$ and Es$_p^{m}(at)$. Hereby $m$ is the number of 
zeros that can be uniquely related to the Maslov index $n$, and $p$ and $a$ 
are fixed numbers that depend on the potential. We find that the 
trigonometric expansions of Ec$_p^m(z)$ and Es$_p^m(z)$ given in Erd\'elyi 
{\it et al.}\ reproduce our numerical results to a high degree of accuracy 
(see Ref.\ \cite{mbmm} for details). 

\bs

\noindent
$^1$see, e.g., {\it Higher Transcendental Functions, Vol.\ III,} A. 
Erd\'elyi {\it et al.}, eds.\ (McGraw-Hill, New York, 1955), Chapter XV.

\bs
\bs


\begin{thebibliography}{31}

\setlength{\itemsep}{-0.25ex}

\bibitem{gutz} M. C. Gutzwiller, J. Math.\ Phys.\ {\bf 12}, 343 (1971).

\bibitem{gubu} M. C. Gutzwiller: {\it Chaos in classical and quantum
               mechanics} (Springer, New York, 1990).

\bibitem{stru} V. M. Strutinsky, Nukleonika (Poland) {\bf 20}, 679
               (1975); V. M. Strutinsky and A. G. Magner, Sov.\ J.
               Part.\ Nucl.\ {\bf 7}, 138 (1976); V. M. Strutinsky, A.
               G. Magner, S. R. Ofengenden, and T. D{\o}ssing, Z.
               Phys.\ {\bf A 283}, 269 (1977).

\bibitem{hh1}  M. Brack, R. K. Bhaduri, J. Law, and M. V. N. Murthy,
               Phys.\ Rev.\ Lett.\ {\bf 70}, 568 (1993); M. Brack, R.
               K. Bhaduri, J. Law, Ch. Maier, and M. V. N. Murthy,
               Chaos {\bf 5}, 317 (1995); {\it ibid.} (Erratum) {\bf
               5}, 707 (1995).

\bibitem{hhun} M. Brack, P. Meier, and K. Tanaka, J. Phys.\ {\bf A 32},
               331 (1999).

\bibitem{qdot} S. M. Reimann, M. Persson, P. E. Lindelof, and M. Brack,
               Z. Phys.\ {\bf B 101}, 377 (1996).

\bibitem{jo}   J. Blaschke and M. Brack, Europhys.\ Lett.\ {\bf 50},
               294 (2000).

\bibitem{fiss} M. Brack, S. M. Reimann, and M. Sieber, Phys.\ Rev.\
               Lett.\ {\bf 79}, 1817 (1997); M. Brack, P. Meier, S. M.
               Reimann, and M. Sieber, in: {\it Similarities and
               differences between atomic nuclei and clusters}, eds.\
               Y. Abe {\it et al.} (A.I.P., 1998) p.\ 17; M. Brack, 
               M. Sieber, and S. M. Reimann, in: {\it Quantum Chaos Y2K}, 
               Proceedings of Nobel Symposium 116, K.-F. Berggren and S. 
               {\AA}berg (Eds.), Physica Scripta Vol.\ {\bf T90} (2001), 
               146. First steps towards the inclusion of spin-orbit 
               interactions are reported in M. Brack and Ch. Amann, in:
               {\it Fission Dynamics of Atomic Clusters and Nuclei}, 
               eds. D. Brink {\it et al.} (World Scientific Publishing,
               2001), in print. 

\bibitem{book} M. Brack and R. K. Bhaduri: {\it Semiclassical Physics},
               Frontiers in Physics, Vol.\ 96 (Addison-Wesley, Reading,
               USA, 1997).

\bibitem{hh}   M. H\'{e}non and C. Heiles, Astr.\ J. {\bf 69}, 73
               (1964).

\bibitem{kabr} K. Tanaka and M. Brack, to be published.

\bibitem{chur} R. C. Churchill, G. Pecelli, and D. L. Rod, in {\it
               Stochastic Behavior in Classical and Quantum Hamiltonian
               Systems}, ed. by G. Casati and J. Ford (Springer-Verlag,
               N.Y., 1979) p.\ 76.

\bibitem{hhdb} K. T. R. Davies, T. E. Huston, and M. Baranger, Chaos
               {\bf 2}, 215 (1992).

\bibitem{vioz} W. M. Vieira and A. M. Ozorio de Almeida, Physica {\bf D
               90}, 9 (1996).

\bibitem{ssun} M. Sieber, J. Phys.\ {\bf A 29}, 4715 (1996); H.
               Schomerus and M. Sieber, J. Phys.\ {\bf A 30}, 4537
               (1997); M. Sieber and H. Schomerus, J. Phys.\ {\bf A
               31}, 165 (1998).

\bibitem{mawu} J. Main and G. Wunner, Phys.\ Rev.\ {\bf A 55}, 1743
               (1997).

\bibitem{scho} H. Schomerus, Europhys.\ Lett.\ {\bf 38}, 423 (1997); J.
               Phys.\ {\bf A 31}, 4167 (1998).

\bibitem{elli} A. Magner, S. N. Fedotkin, K. Arita, T. Misu, K.
               Matsuyanagi, T. Schachner, and M. Brack, Prog.\ Theor.\
               Phys.\ (Japan) {\bf 102}, 551 (1999).

\bibitem{safe} A. Magner, S. N. Fedotkin, and M. Brack, work in
               progress.

\bibitem{feig} M. J. Feigenbaum, J. Stat.\ Phys.\ {\bf 19}, 25 (1978);
               see also M. J. Feigenbaum, Physica {\bf 7 D}, 16 (1983).

\bibitem{fei2} T. C. Bountis, Physica {\bf 3 D}, 577 (1981).

\bibitem{fei3} J. M. Greene, R. S. McKay, F. Vivaldi, and M. J.
               Feigenbaum, Physica {\bf 3 D}, 468 (1981).

\bibitem{nong} The non-generic nature of the bifurcations in potentials
               with discrete symmetries has been discussed, e.g., by M.
               A. M. de Aguiar, C. P. Malta, M. Baranger, and K. T. R.
               Davies, Ann.\ Phys.\ (N.Y.) {\bf 180}, 167 (1987), and
               by Mao and Delos \cite{maod}.

\bibitem{mbmm} S. Fedotkin, M. Mehta, K. Tanaka, and M. Brack, to be 
               published.

\bibitem{hill} G. W. Hill, Acta Math.\ {\bf 8}, 1 (1886).

\bibitem{mawi} W. Magnus and S. Winkler: {\it Hill's Equation}
               (Interscience Publ., New York, 1966).

\bibitem{edmo} A. R. Edmonds, J. Phys.\ {\bf A 22}, L673 (1989).

\bibitem{maod} J.-M. Mao and J. B. Delos, Phys.\ Rev.\ {\bf A 45}, 1746
               (1992).

\bibitem{w1}   B. Barbanis, Astr.\ J. {\bf 71}, 415 (1966).

\bibitem{pert} M. Brack, S. C. Creagh, and J. Law, Phys.\ Rev.\ {\bf A
               57}, 788 (1998).

\bibitem{main} J. Main, G. Wiebusch, A. Holle, and K. H. Welge, Phys.\
               Rev.\ Lett.\ {\bf 57}, 2789 (1986).

\bibitem{sum}  M. Y. Sumetskii, Sov.\ Phys.\ JETP {\bf 56}, 959 (1983).

\bibitem{mai2} J. Main, G. Wiebusch, A. Holle, and K. H. Welge, Z.
               Phys.\ {\bf D 6}, 295 (1987).

\end{thebibliography}
\end{document}